\documentclass[12pt]{article}

\usepackage{graphicx}
\usepackage{epstopdf}
\usepackage{amsmath}
\usepackage{amsfonts}
\usepackage{amssymb}
\usepackage{enumerate}
\usepackage{amsthm}
\usepackage{array}
\usepackage{epsfig}
\usepackage[small,bf]{caption}
\usepackage{appendix}
\usepackage{dsfont}
\usepackage{hyperref}
\def\1ad{\mbox{\normalsize $^1$}}
\def\2ad{\mbox{\normalsize $^2$}}
\def\3ad{\mbox{\normalsize $^3$}}
\def\4ad{\mbox{\normalsize $^4$}}
\def\5ad{\mbox{\normalsize $^5$}}
\def\6ad{\mbox{\normalsize $^6$}}
\def\7ad{\mbox{\normalsize $^7$}}
\def\8ad{\mbox{\normalsize $^8$}}

\def\beq{\begin{equation}}                     % 
\def\eeq{\end{equation}}                       %
\def\bea{\begin{eqnarray}}                     %         %
\def\eea{\end{eqnarray}}                       %       % 
\def\dj{\hbox{d\kern-0.347em \vrule width 0.3em height 1.252ex depth
-1.21ex \kern 0.051em}}

\def\half{{1\over 2}\,}

\def\pt{\partial}

\def\ord{{\cal O}}

\def\shalf{{\mbox{$\half$}}}

\def\sthreeovertwo{{\mbox{${3\over 2}$}}}

\def\Dirac{\,\raise.15ex\hbox{/}\mkern-13.5mu D}
\def\dirac{\,\raise.15ex\hbox{/}\kern-.57em \partial}
\def\pslash{\,\raise.15ex\hbox{/}\kern-.57em p}

\begin{document}

                     %
%%%%%%%%%%%%%%%%%%%%%%%%%%%%%%%%%%%%%%%%%%%%%%%%
%%%%%%%%%%%%%%%%%%%%%%%%%%%%%%%%%%%%%%%%%%%%%%%%%%%%%%%%%%%%%%%%%%%%%%%%% e
%%%%%%%%%%%% text begins %%%%%%%%%%%
%%%%%%%%%%%%%%%%%%%%%%%%%%%%%%%%%%%%%%%%%%%%%%%%%%%%%%%%%%%%%%%%%%%%%%%%%%

\newcommand{\sheptitle} {Spontaneous fragmentation of topological black holes} \newcommand{\shepauthora} {{\sc Jos\'e
    L.F.~Barb\'on and Javier Mart\'{\i}nez--Mag\'an}}

\newcommand{\shepaddressa} {\sl
  Instituto de F\'{\i}sica Te\'orica  IFT UAM/CSIC \\
  Facultad de Ciencias C-XVI \\
  C.U. Cantoblanco, E-28049 Madrid, Spain\\
  {\tt jose.barbon@uam.es}, {\tt javier.martinez@uam.es} }

\newcommand{\shepabstract} {
  \noindent

 We study the  metastability   of Anti-de Sitter topological black holes with compact hyperbolic horizons. We focus on the five-dimensional case, an AdS/CFT dual to thermal states in the maximally supersymmetric large-$N$ Yang--Mills theory, quantized on a three-dimensional compact hyperboloid. 
 We estimate the various rates  for quantum-statistical D3-brane emission, using WKB methods in the probe-brane approximation,  including thermal tunneling and Schwinger pair production. The topological black holes are found to be metastable at high temperature. At low temperatures, D-branes are emitted without exponential suppression  in superradiant modes,  producing an instability  in qualitative agreement with expectations from weakly-coupled gauge dynamics. }

\begin{titlepage}
  \begin{flushright}
    % \today
    {IFT-UAM/CSIC-10-27\\
      % {\tt hep-th/yymmnnn}}
    }

\end{flushright}
\vspace{0.5in} \vspace{0.5in}
\begin{center} {\large{\bf \sheptitle}}
  \bigskip\bigskip \\ \shepauthora \\ \mbox{} \\ {\it \shepaddressa} \\
  \vspace{0.2in}
  % \vspace{1in}
  % \bigskip\bigskip  \shepauthorb \\ \mbox{} \\ {\it \shepaddressb} \\
  % \vspace{0.5in} \vspace{1in}

  {\bf Abstract} \bigskip \end{center} \setcounter{page}{0} \shepabstract
% \vspace{1in}
\vspace{2.7in}
\begin{flushleft}
  % CERN-TH/2004-059\\
  % March 2001
  \today
\end{flushleft}

%%%%%%%%%%%%%%%%%
%% \hfill\jobname
%%%%%%%%%%%%%%%%%

\end{titlepage}

\newpage

%%%%%%%%%%%%%%%%%%%%%%%%%%%%%%%%%%%%%%%%%%%%%%%%%%%%%%%%%%%%%%%%%%%%%%

\setcounter{equation}{0}

\section{\label{intro} Introduction}

\noindent

Topological black holes may be counted among the most  exotic specimens   in the quite vast  bestiary of black objects studied recently.
They are characterized by event horizons of complicated topology, obtained by modding a hyperbolic hyperplane by a freely acting discrete  isometry, and exist in an asymptotically Anti-de Sitter spacetime  (AdS) with compatible topological identifications  \cite{toporefs, roberto, topoantiguo}. Similar black objects 
arise naturally in string theory as the near-horizon limit
of thermally excited D-branes with such world-volume topology. In particular, for the case of D3-branes one finds topological black holes in AdS$_5$ with a direct interpretation in terms of
a dual four-dimensional conformal field theory (CFT) living on a compact 3-hyperboloid $\Sigma_3={\bf H}^3 /\Gamma$, where $\Gamma$ is a freely acting discrete isometry. 

AdS/CFT for CFTs on spaces of negative curvature has been comparatively less studied, since
most physical questions investigated so far can be appropriately formulated in more standard examples of CFTs defined on flat tori or spheres. There are, however, good reasons to study these models in various contexts.

The low-lying spectrum of field theories on spaces of the form
$\Sigma_n = {\bf H}^n /\Gamma$ is quite interesting. While the spectral gap of the Laplacian operator on $\Sigma_n$ is controlled by the curvature radius $\ell$, the volume of the
compact hyperboloid is determined by the overall size  induced by the $\Gamma$ identification. This means that Kaluza--Klein models based on compact hyperboloids can support
hierarchies of couplings with good decoupling properties of KK modes \cite{trodden}. When such compactification manifolds are contemplated in string theory, new phenomena take place regarding the dynamics of winding modes, beyond the usual rules of T-duality (cf. \cite{evauno}). 

The effect of negative curvature on  the stability of CFT's and their AdS/CFT counterparts has been studied from various points of view (see for example  \cite{frag, seiwit, buchel, rabadan}). Any scalar degree of freedom with a conformal coupling has a background curvature coupling of the form (in four dimensions) $-\phi^2 R / 6$, where $R$ is the Ricci scalar. This term 
renormalizes the  $\phi$ mass increasing its stability for positive background curvature, i.e. the case of the sphere for instance. Conversely, for the theory on $\Sigma_n$ this term induces a tachyonic shift in the effective mass-squared of $\phi$. Since this shift is of order
$-1/\ell^2$, the same order of magnitude as the spectral gap, only a few modes can be tachyonic
in practice, sometimes just the zero mode.  Our main concern in this note is the study of whether this instability of the zero mode may be converted into {\it metastability} by thermal effects. 

AdS/CFT models for CFT's on hyperboloids have been motivated recently as an interesting toy model for the holographic study  of black hole interiors \cite{horo}. 
 By a clever use of different conformal frames, a  specific topological black hole on AdS$_5$ can be related to a four-dimensional CFT on a Milne-type cosmology with compact hyperbolic spatial sections. Dynamical processes in which  black holes are formed by sending D-branes from infinity
may be studied in the particular case of the maximally supersymmetric Super Yang--Mills (SYM) theory with $SU(N)$ gauge group.  One sets up the  scattering of an initial state with large values of the
adjoint Higgs fields into an intermediate resonance of vanishing Higgs fields \cite{eva, horo}, whose dynamics appears related to the crunch singularity in the black hole interior. The black hole corresponds to the thermalization of these Higgs fields near the origin of field space, and the subsequent decay
of the black hole would complete the S-matrix description of the process. 

Our main interest in this note is the  decay phase, whereupon the quasi-static black hole
ejects the branes back to infinity. 
We concentrate on the $n=3$ case, i.e. topological black holes in AdS$_5$ with a dual description in terms of thermal states of  the maximally supersymmetric Yang--Mills theory in four dimensions (see  \cite{buchel} for
 earlier work on this system.) Our results should  generalize naturally to all  $n>1$, provided a microscopic description of the system is available (such as the simple cases based on M2 or M5 branes). The $n=0, 1$ cases pose additional  subtleties, even for the theories defined on spatial manifolds of positive curvature, and   we shall not discuss them here (see \cite{frag, seiwit}).

\section{Topological black holes}

\noindent

Let us consider a family of black-hole spacetimes  with hyperbolic horizons and AdS$_5$ asymptotic  metric, 
\begin{equation}\label{bhmet}
ds^{2}=-f\left( r\right) dt^{2}+\frac{dr^{2}}{f\left( r\right)}+\frac{r^{2}}{\ell ^{2}}d\Sigma_3^{2}\;,
\end{equation}
where
\begin{equation}
\label{f}
f(r)=\frac{r^{2}}{\ell ^{2}} -1  -\frac{\mu}{r^{2}}
\end{equation}
and $d\Sigma_3^{2}$ is the metric of a three-dimensional compact hyperbolic space ${\bf H}^{3} /\Gamma $ of curvature radius $\ell$  and volume $V$.  In what follows we shall use units in which the AdS radius of curvature is set to unity,  $\ell =1$. 
In these units, the   minimum value of $\mu$ compatible with regularity is $\mu_{\rm min} = -1/4$, which defines the extremal black hole, whereas the $\mu=0$ case  has the special property of being locally isometric to pure AdS. The associated Hawking temperature is $T = \beta^{-1}$ with
\beq\label{invhaw}
\beta = {2\pi r_0 \over 2r_0^2 -1}\;,
\eeq
and 
\beq\label{horrad}
r_0^2 = \shalf\left(1+\sqrt{1+4\mu}\right)
\eeq
determines  the horizon radius. The energy of these black holes reads
\beq\label{masa}
M= {3V \over 16\pi G} \left(\mu + {1 \over 4} \right) = {3V \over 16\pi G} \left({1\over 4} + r_0^4 - r_0^2 \right)\;,
\eeq
whereas the entropy takes the usual form
\beq\label{entropy}
S= {V \over 4G} \,r_0^3\;, 
\eeq
with $G$ denoting the effective five-dimensional Newton constant. 
Using the standard dictionary to parameters of the dual $SU(N)$ gauge theory we have (cf. \cite{roberto}) 
\beq\label{energyYM}
\begin{split} M(N,T) & = {3\pi^2 N^2   \over 32}\,a(T)^2\,V \,T^4 \;, \\ 
  S(N,T) &= {\pi^2 N^2 \over 16}\, 
a(T)^3\,V \,T^3\, \;,
\end{split}
\eeq
where we denote 
$$
a(T) \equiv 1+ \sqrt{1+{2\over (\pi T)^2}}\;.
$$
The free energy $F = M-TS$ is then given by
\beq\label{freenn}
F(N,T)= - T\,S(N,T)\,\left(1-{3\over2 a(T)}\right)\;,
\eeq
and the associated chemical potential for the Ramond--Ramond (RR) charge
\beq\label{chem}
\mu_N \equiv {\pt F \over \pt N} = -{1\over N} T\,S(N,T)\, \left(2-{3\over a(T)}\right)
\eeq
is negative at all temperatures and vanishes in the zero-temperature limit. In this form, the thermodynamic functions refer to thermal states of the $SU(N)$, maximally supersymmetric Yang--Mills theory quantized on $\Sigma_3$, in the limit of large  $N$ and large 't Hooft coupling, $N \gg \lambda \gg 1$, where  $\lambda \equiv g_{\rm YM}^2 N$.

The thermodynamics implied by these expressions is standard in the large temperature limit, but it is quite peculiar for very low temperatures. As pointed out in \cite{roberto}, the  zero-energy configuration, corresponding to the $T=0$ black hole, retains a macroscopic entropy of order $N^2$. This seems to be a radical effect of the strong coupling limit,  implicit in the supergravity description. 

  The specific heat of these black holes is positive, and so they can come to equilibrium with radiation trapped in AdS. This means that these black holes are stable with respect to radiative decays through particle degrees of freedom in the
supergravity multiplet. Again, the positivity of the specific heat persists down to zero temperature, so that there are no perturbative signs of any instabilities for the highly degenerate zero-energy level.

The black holes themselves can be viewed as thermal states in the gauge theory with all the adjoint Higgs fields concentrated around $\phi=0$. Such a configuration should  suffer from instabilities because of the classical
tachyonic potential $V(\phi)_{\rm cl} \propto -\phi^2$ of the Higgs zero mode on $\Sigma_3$. Still, the Higgs field can be locally  trapped by thermal effects near the origin of field space provided the temperature is large enough, i.e. for $T\gg 1$, in units of the curvature radius of  $\Sigma_3$, the
effective potential obtained by integrating out non-zero modes in a thermal state around $\phi=0$ should contain a standard thermal mass of the form $m^2_\beta \sim \lambda\, T^2$, which overcomes the tachyonic term induced from the background curvature for $\beta \ll \sqrt{\lambda}$, whereas at large field strengths the supersymmetric nature of the high-energy theory should yield temperature-dependent effects in the quantum effective potential suppressed by powers of $T/\phi$. Under these circumstances we can expect the classical negative quadratic term to dominate at large $\phi$, thus depicting an unbounded effective potential with a local minimum at $\phi=0$ .  For very low temperatures  the tachyonic mass dominates even in the region of small fields, and we expect the dynamics  to be described by a classical runaway of the Higgs fields down the
tachyonic potential. 

 A full calculation of this effective potential to leading order in perturbation theory in the SYM theory is an interesting open problem. In this note we shall perform a calculation motivated by the strong-coupling description of the system as a topological black hole on AdS, by considering the fragmentation of the black hole by D3-brane emission. We shall work in the brane-probe approximation and consider the Hawking radiation of a single D3-brane, with the subsequent quantum-statistical process
of decay into the asymptotic runaway region. In the SYM interpretation, we are then considering the spontaneous decay of
the thermal state at the origin of field space, via the symmetry breaking pattern $SU(N) \rightarrow SU(N-1) \times U(1)$.

\section{Brane probes}

\noindent

In the brane picture of the breaking pattern $SU(N) \rightarrow SU(N-1) \times U(1)$,  the probe-brane effective action is identified with the result of integrating out all the $SU(N-1)$ degrees of freedom, and the corresponding thermal effects incorporated into the temperature dependence of the effective action.  In turn, this temperature dependence enters the probe-brane calculation via the $\mu$ parameter in the black hole metric (see for instance  \cite{kiritsis}). In our analysis, we will obviate the dynamics with respect to the R-symmetry group, i.e. the motion of the D3-branes in the `internal'  ${\bf S}^5$, so that we only consider the rigid motion of D3-branes in AdS$_5$. We also freeze to their vacuum values the excitations of the world-volume gauge fields in our probe brane. 

Under such conditions, we  take the world-volume $\cal{W}$ of the probe D3-brane to be the embedding into 
AdS$_5 \times {\bf S}^5$   of ${\bf R} \times \Sigma_3  \times P$, where $P$ is a fixed point on ${\bf S}^5$,  ${\bf R}$ represents time and the whole $\Sigma_3$ is mapped rigidly at the same radial position in the coordinates of (\ref{bhmet}), given by a function $r(t)\geq r_0$.  The probe effective action takes the form
\beq\label{dprobe}
I = -{\cal T}_{{\rm D}3}\,\int_{\cal W} \sqrt{-{\rm det}(h_{ab})} + \varepsilon\, {\cal T}_{{\rm D3}} \int_{\cal W} C_4 \;,
\eeq
where $h_{ab}$ is the induced metric on the world-volume and $C_4$ is the  RR four-form coupling minimally to the D3-brane.  D3-branes correspond to $\varepsilon =+1$, whereas $\varepsilon = -1$ yields the action of an antiD3-brane.  

 A natural additive normalization for $C_4$ is obtained by requiring it to be well-defined on the Euclidean black-hole spacetime, with topology ${\bf R}^2 \times \Sigma_3 \times {\bf S}^5$, with the Euclidean time being an angular variable in the ${\bf R}^2$ factor. In this case the projection of $(1/4)dC_4$ onto the AdS factor   is the volume form of the Euclidean AdS black-hole,  and the Euclidean action takes the Wess--Zumino form
\beq\label{edpro}
I_E = {\cal T}_{{\rm D}3} {\rm Vol} \left[\,{\cal W}\,\right] - 4\,\varepsilon \,{\cal T}_{{\rm D}3} {\rm Vol}\left[\,{\overline{\cal W}}\,\right]\;,
\eeq
with ${\overline{\cal W}}$ the compact four-dimensional manifold having ${\cal W}$ as its boundary.  This prescription demands that the AdS projection of the RR form be given by 
\beq\label{presc}
C_4\big|_{\rm AdS} = (r^4 - r_0^4) \,d\Sigma_3 \wedge dt 
\;,
\eeq 
so that it vanishes at the  horizon in the static frame. 

  Evaluating the action for the particular embedding under consideration we find, in the real-time static coordinate system 
\beq\label{probecoor}
I = -m \int dt \left[ r^3 \sqrt{f(r) - {{\dot r}^2 \over f(r)}} - \varepsilon \,(r^4 -r_0^4) \right]\;,
\eeq
where ${\dot r} \equiv dr/dt$ and we introduce an `effective mass parameter' 
\beq\label{mpara}
m\equiv  V \,{\cal T}_{{\rm D}3}   = {N V \over  2\pi^2}\;,
\eeq
 in units $\ell=1$.

We can obtain a first indication of the physics implied by (\ref{probecoor}) by looking at the slow motion regime 
\beq\label{slow}
L = \shalf m \,{\dot  \varphi}^2 - V_{s} ( \varphi) + {\cal O}({\dot \varphi}^4)\;,
\eeq
with $\varphi$ another radial variable related to $r$ by\footnote{See \cite{seiwit} for the general form of such field redefinitions.} $d\varphi =dr \,r^{3/2} f(r)^{-3/4} $, matching to the  zero-mode of the Higgs field in the dual SYM theory by the relation $
m \varphi^2 = V \phi^2$.  The resulting effective static potential takes the form 
\beq\label{nrel}
V_{s} (\varphi) = m\left[r^3 \sqrt{f(r)} + \varepsilon(r_0^4  -r^4)  \right]\;,
\eeq
where the functional dependence $r(\varphi)$ is implicitly understood.  Asymptotically, as $r\rightarrow \infty$
we have $\varphi \propto r$ and the static potential for branes ($\varepsilon = 1$)  approaches the tachyonic regime  
$V_{s} (r)\rightarrow - m\, r^2 /2$. We thus 
 recover the leading tachyonic potential at large values of the Higgs field, the powers of $r^4 \propto \varphi^4 $ canceling out because of the BPS property of the D3-branes.  Conversely, for antibranes ($\varepsilon =-1$) the potential shows a quartic rise at large radius. 

For generic values of $r_0$, the function $V_{s} (r)$ is tangent to a vertical line at the horizon,  indicating that the naive low-velocity approximation breaks down in the near-horizon region. For this reason, we shall not study the detailed near-horizon dynamics in terms of (\ref{slow}). Despite this fact, it will be shown that some properties of the motion are correctly captured by the static potential, such as the fact that neither branes nor antibranes can propagate with negative energy in the vicinity of the horizon. Branes of arbitrarily negative energy can exist, but their motion has a turning point `on the other side of the barrier' (cf. figure 1). 

\begin{figure}
  \label{potestatico}
  \begin{center}
    \epsfig{file=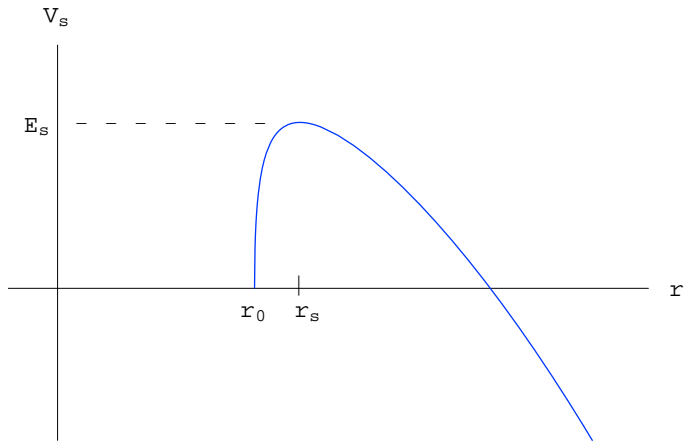, width= 9.5cm}
   \caption{ \small Picture of the static potential $V_s (r)$ for probe D-branes, showing the local maximum at the `sphaleron' point $r_s$. Near-horizon D-branes  with  energy within the interval $0 = V_s (r_0) < E < V_s (r_s)= E_s$ must tunnel through the barrier in order to escape, whereas D-branes with energy above $E_s$ run away classically.  }
  \end{center}
\end{figure}

Another exact property of the static potential turns out to be the location of the maximum, $r_s$, and the corresponding `sphaleron energy' $E_s \equiv V_s (r_s)$, characterizing the minimum energy for which the asymptotic motion has no turning points above the horizon. The sphaleron energy is related to the location of the maximum by the equation
$$
E_s = \half m \left(r_0^4 + r_0^2 - {3\over 2} r_s^2 \right)\;.
$$
In the high-temperature limit, $r_0 \gg 1$, the barrier widens as $r_s \rightarrow (r_0^8 /2)^{1/6} $ and its height  grows as $E_s \rightarrow \shalf m r_0^4$. In the low-temperature limit, the barrier turns off as $r_s \rightarrow r_0 \rightarrow 1/\sqrt{2}$ and
$E_s \rightarrow 0$. In particular, the static potential at zero temperature, $V_s (r) \big |_{T=0} = \shalf m (\shalf -r^2)$, is monotonically decreasing for all radii above the horizon, showing that the barrier has all but disappeared at $T=0$. This fact suggests that perhaps  the barrier is not efficient in containing the branes for sufficiently low temperatures. 

The `permeability' of the barrier depends in practice on the typical energy of the branes impinging on it. At weak coupling, this is dictated by the thermal distribution of the scalar field configurations in the SYM theory. At strong coupling, the static potential depicted in figure 1 does not extend by itself to the origin of field space. Instead, we must view the horizon as providing the thermal-state initial condition  for the motion of the branes
in the region $r\geq r_0$. 
Hence, we will adopt the physical prescription that
D-branes are emitted by the horizon with a Hawking spectrum. This means that the rate of `fragmentation' in the pattern $SU(N) \rightarrow SU(N-1) \times U(1)$ must be computed as the convolution of  the Hawking rate with  the
`grey-body' factor resulting from the quantum barrier penetration. 

While these comments serve to frame the spirit of our computation, a number of remarks are in order regarding the detailed implementation. First, the nonrelativistic approximation implicit in $V_s (r)$ breaks down in the vicinity of the horizon, so that the WKB approximation to the decay rate must be implemented at the relativistic level. Second, the height of the
static barrier is of order $E_s =\ord(mr_0^4)$, which coincides in order of magnitude with the size of the chemical potential $|\mu_N |$ (except at very low temperatures). This implies that the D-branes are emitted with a typical energy of order $E_s$, calling into question the efficiency of the barrier in containing the decay. In other words, the metastable character of the black hole is potentially sensitive to the leading finite-$N$ effects on the energetic balance of the emission process.  Finally, the height of the barrier is proportional to the brane mass parameter, $m$, so that relativistic effects associated to pair-production compete with the standard tunneling processes and deserve specific study. 
In what follows we address each of these questions separately, keeping in mind the basic physical picture suggested by figure 1.

\noindent

\subsection{Non-linear dynamics}

\noindent

We now describe the qualitative properties of the brane motion in the {\it exterior} of the topological black holes, i.e. for radial coordinates $r\geq r_0$, including all nonlinear effects implied by the Lagrangian (\ref{probecoor}). The conserved canonical energy 
\beq\label{energy}
E = {\dot r}\,{\partial L \over \partial {\dot r}} - L
\;,
\eeq
satisfies 
\beq\label{truesign}
{E \over m} + \varepsilon \,(r^4 -r_0^4)  = {r^3 f(r) \over \sqrt{f(r) - {\dot r}^2 /f(r)}}\;.
\eeq
Squaring this equation and solving for ${\dot r}$ we find  a conveniently intuitive  description of the dynamics as a zero-energy motion in the effective non-relativistic problem 
\beq\label{effpot}
{\dot r}^2 + V_{\rm eff} (r) =0\;,
\eeq
where the effective potential is given by
\beq\label{effpotexp}
V_{\rm eff} (r) = f(r)^2 \; \left[ {r^6 f(r) \over \left({E \over m} + \varepsilon \,(r^4 -r_0^4) \right)^2} -1 \right]\,.
\eeq 
Since this potential problem  is obtained by squaring  (\ref{energy}),  some information about signs is lost in the process, and must be recovered from (\ref{truesign}). 
Positiveness of the right-hand side of (\ref{truesign})  implies that  necessarily $r^4 > r_0^4 -E/m$ for branes and $r^4 < r_0^4 + E/m$ for antibranes. Therefore, the potential $V_{\rm eff} (r)$ with $\varepsilon =1$  describes at the same time  branes of energy $E$ moving `to the right of the pole' and antibranes of energy $ -E$ moving `to the left of the pole'.  

\begin{figure}
  \label{barreras}
  \begin{center}
    \epsfig{file=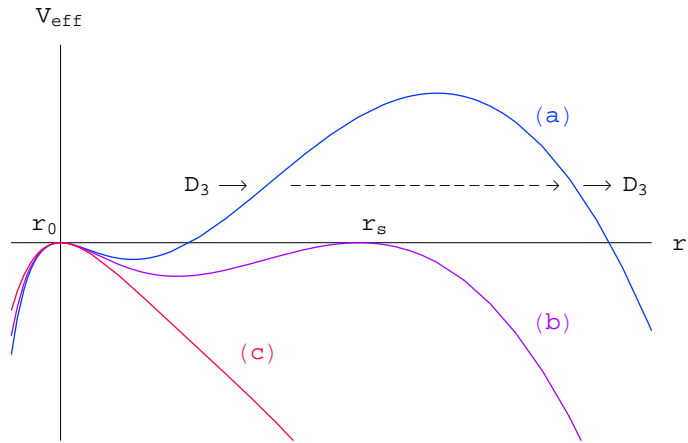, width= 9.5cm}
   \caption{ \small A picture of the non-linear effective potential just above  the threshold of D-brane radiation. Curve (a) corresponds to $0<E< E_s$, with curve (b) giving the {\it sphaleron} case, $E = E_s$. Finally, for  case (c), with $E>E_s$, there is no barrier, and the potential becomes monotonically decreasing for $E \gg E_s$ at all radii outside the horizon.  In case (a) D3-branes emitted thermally from  the horizon at $r=r_0$ can tunnel across the barrier and subsequently run away to the asymptotic region. As $T\rightarrow 0$, we have $E_s \rightarrow 0$ and the barrier tends to disappear.}
  \end{center}
\end{figure}

With no loss of generality, we then concentrate on the case $\varepsilon = 1$. Expanding the potential at large radius 
we find the universal asymptotic behavior  $V_{\rm eff} (r\rightarrow \infty) \rightarrow - r^2$ as expected. On the other hand, for large and negative values of $\omega$ we are always guaranteed a positive pole $r_p = (r_0^4 -E/m)^{1/4}$ at large radius, showing that $V_{\rm eff} $ must have a zero
at large radius of order $r_+ > r_p$, the turning point for the runaway trajectories.

 As advanced in the previous section, we have two qualitatively different situations depending on whether the canonical energy of the probe brane is positive or negative. For $E< 0$
the pole is in the physical region and only antiD-branes can propagate in the immediate vicinity of the horizon. Conversely, for $E > 0$ the pole is formally `inside' the black hole and thus no antiD-branes can propagate outside the horizon, whereas D-branes can actually fall into  the black hole, if found in its vicinity.

 Away from extremality, that is to say for $\mu > -1/4$, the profile function   $f(r)$ is positive for  all $r>r_0$ and has a simple zero
at $r=r_0$. Therefore, both the potential and the first derivative vanish at the horizon, with a negative value of the second derivative, for all $E \neq 0$. It follows that the effective potential starts with a gentle quartic descent in the vicinity of the horizon, which is a region of allowed classical motion for either antibranes ($E<0$) or branes ($E > 0$). The marginal situation $E =0$ has the pole cancelled out by the zero of $f(r)$ so that neither branes nor antibranes can propagate near $r_0$ in this case, except for the extremal black hole, $\mu=-1/4$, on which branes can still fall.

 \begin{figure}
  \label{polo}
  \begin{center}
    \epsfig{file=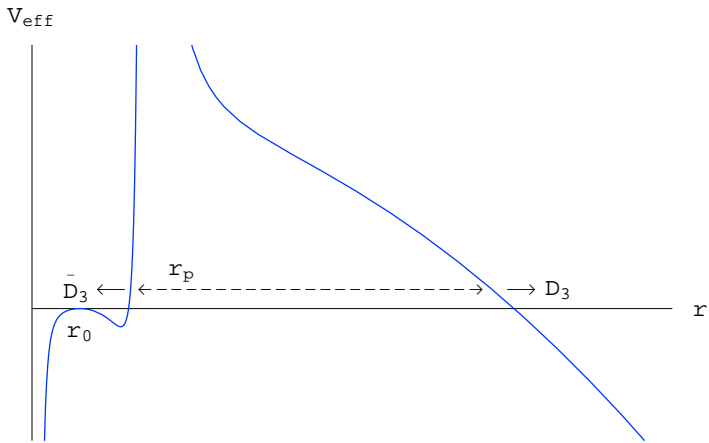, width= 9.5cm}
   \caption{ \small The effective potential $V_{\rm eff}$ below threshold, for $E <0$, has a pole at $r_p > r_0$. D-branes can only propagate as zero-energy motions to the right of the second turning point, whereas antibranes  with energy ${\overline E} = -E >0$ can be trapped between the horizon and the first turning point, to the left of the pole. The analog of the Schwinger pair production feeds such brane-antibrane pairs at the expense of the background RR field.}
  \end{center}
\end{figure}

For $E \gg 0$ the potential is monotonically decreasing for $r\geq r_0$. However,
near the pole threshold, within an interval $0 <E< E_s $ a finite barrier develops in the non-extremal case ($\mu > -1/4$). At the `sphaleron'  energy, $ E_s $,  there is a local maximum at $r_s > r_0$ with exactly vanishing potential $V'_{\rm eff} (r_s) =V_{\rm eff} (r_s) =0$, so that $r(t) = r_s$ is an unstable static trajectory. As $E$ goes
slightly below $E_s$ the local maximum becomes positive and the barrier develops until
 the pole gets superimposed on the barrier for $E < 0$. One can explicitly check that this definition of the sphaleron point coincides with the one given above in terms of the static potential.  Finally, as $E \ll 0$ the pole at $r_p \sim (r_0^4 -E / m)^{1/4}$  migrates far away from the horizon, accompanied by zeros of the effective potential $V_{\rm eff} (r_\pm) =0$, with $r_- = {\cal O}(r_p)$ and $r_+ = {\cal O}(r_p^2)$.

\subsection{Energy conventions and Hawking spectrum}

\noindent

As mentioned in the previous section, the marginal height of the barrier, compared to the typical energy of
D3-branes, requires a  careful treatment of the different sources of energy differences in the emission process.
We shall assume  the physical boundary condition at the horizon that  D-branes are emitted by the black hole at the appropriate Hawking rate.  Lacking a complete theory of Hawking radiation for D-brane objects,\footnote{It might be possible to develop a formalism based on the ideas of \cite{wil}.} we will adopt here a  physical prescription, demanding that the decay rate  be proportional to the
 ratio of density of states before and after the emission of a D-brane, 
\beq\label{ratio}
\Gamma_H \propto \exp\left(\Delta S\right)\;, 
\eeq
where $\Delta S = S_f - S_i$ for the initial and final black hole states.  
 If we emit a single D3-brane
of  energy $\omega$, we have
$\Delta S = S(N-1, M-\omega) - S(N, M)$. Using (\ref{energyYM})  we find, to leading order in the $1/N$ expansion, the expected exponential rate 
\beq\label{rate}
\Gamma_H (\omega) \propto \exp\left(-\beta\, \omega + \beta \mu_N\right)
\;,
\eeq
where $\mu_N$ is given by (\ref{chem}). 
 We can also consider the emission of antiD-branes, in which case
the chemical potential has the opposite sign, $\mu_{\bar N} = - \mu_N$. Notice that $\mu_N$ is negative, suppressing the emission of D-branes and enhancing the emission of antiD-branes. 

We shall apply the relation (\ref{rate}) only when it is semiclassical, i.e. $\Gamma_H \ll1$. When the energy and temperature parameters are taken to extremes, so that the exponential approximation (\ref{rate}) gives $\exp(-\beta \omega + \beta \mu_N) \sim 1$, we  assume  the rate to be controlled, in order of magnitude, by the temperature: $\Gamma_H \sim T$, up to a dimensionless  ${\cal O}(1)$ function of $T\ell$.   It would be interesting to determine if this function is smooth in the vicinity of the origin, which would ensure a power-law suppression of the decay rate in the extremal $T\rightarrow 0$ limit. Such a study is however beyond the crude semiclassical methods used here, since one can expect subtleties related to the superradiant character of the modes with $\omega<\mu_N$. 

In order to determine the complete decay rate, we need to relate the energy $\omega$, defined in terms of the thermodynamics of the gauge theory, at fixed $V$ and $\ell$, to the canonical energy $E$ defined by equation
(\ref{energy}). Let us rewrite the total energy of the black hole as
\beq\label{transl}
M = {3 V \over 16\pi G} \left( \mu + {1\over 4}\right) = {3N^2 V \over 8\pi^2} \left(\mu + {1\over 4}\right), 
\eeq
where we have used the convention $\ell =1$ in the second equation. On emitting a  D3-brane of energy $\omega$, the black hole loses one unit of RR charge, so that $\Delta N = N_f - N_i = -1$ and 
\beq\label{lead}
-\omega= M_f - M_i  \approx \Delta N\, {\partial M \over \partial N}\Big |_{\mu} + \Delta \mu \,{\partial M \over \partial \mu}\Big |_N =  -{6VN \over 8\pi^2} \left(\mu + {1\over 4}\right) +   {3V \over 16\pi G} \,\Delta\mu \;, 
\eeq
here $\Delta \mu = \mu_f -  \mu_i $ is the difference of ADM mass parameters as the emission takes place, which determines the  ADM mass of the brane: 
\beq\label{admmas}
-\omega_{\rm ADM} = 
{3V \over 16\pi G}\, \Delta \mu\;.
\eeq
Hence, we find the relation
\beq\label{relad}
\omega = \omega_{\rm ADM} +{2\over N} M = \omega_{\rm ADM} +  {3\over 2} m \left(r_0^4 - r_0^2 + {1\over 4}\right)\;.
\eeq
It remains now to relate the ADM mass to the canonical mass, a task that we shall undertake by adapting the results of reference \cite{us}. In order to proceed, we recall that the ADM formalism treats the branes as codimension-two defects in the effective five-dimensional gravitational theory on AdS$_5$, with fixed value of the Newton's constant $G$. On the other hand, after reduction on the ${\bf S}^5$, the RR form $C_4$ gives rise to a five-dimensional cosmological constant with a quantized value, jumping by an amount of ${\cal O}(1/N)$ on crossing a D3-brane.  One can then solve Einstein's equations in the defect approximation, using Israel's junction conditions \cite{israel} (see also \cite{junct, us}) for a metric defined in two AdS patches:  an AdS black hole metric of type (\ref{bhmet})  on the `exterior' of the brane, characterized by a profile function $f_i (r) = r^2 /\ell_i^2 -1 - \mu_i /r^2$,   and an analogous interior metric, with profile function $f_f (r) = r^2 /\ell_f^2 -1 - \mu_f /r^2$.  The two metrics are matched at the world-volume of the D3-brane, with induced metric
\beq\label{induced}
ds^2_{\rm induced} = -d\tau^2 + r(\tau)^2 d\Sigma_3^2\;,
\eeq
with $r(\tau)$ the trajectory of the brane in terms of its proper time.  For any such defect of tension
$\sigma$ the junction conditions imply 
\beq\label{junc}
\left[f_f (r) + \left({dr \over d\tau}\right)^2\right]^{1/2} - \left[f_i (r) + \left({dr \over d\tau}\right)^2\right]^{1/2} = {8\pi G \over 3} \sigma \,r(\tau)\;.
\eeq
The relation between the asymptotic time appearing in (\ref{bhmet}) and the proper time of (\ref{induced}) can be found by matching the exterior metric and the induced metric, with the result
$$
{dt \over d\tau} = \sqrt{{1 \over f(r)} + {1\over f(r)^2}\left({dr\over d\tau}\right)^2 }\;,
$$
where we denote $f_i (r) \equiv f(r)$. Upon  using this relation and squaring (\ref{junc}) twice we can represent the motion in the form 
(\ref{effpot}) with the potential (cf. \cite{us}) 
\beq\label{newpp}
{\widetilde V}_{\rm eff} (r) = f(r)^2 \left[ {r^6 \sigma^2 V^2 f(r) \over (\omega_{\rm ADM} + q\, V\, r^4)^2} -1\right]\;,
\eeq
where the effective charge $q$ is given by
$$
q= {3 \over 16\pi G} \left( {1 \over \ell_f^2} - {1\over \ell_i^2}  \right) - {4\over 3} \pi G \sigma^2\;.
$$
The potential ${\widetilde V}_{\rm eff} (r)$ matches exactly  the $\varepsilon =1$ effective potential $V_{\rm eff} (r)$ under the natural BPS identification 
$q=  \sigma = {\cal T}_{{\rm D}3}$, and the further additive  map  of energy parameters:
\beq\label{admtran}
\omega_{\rm ADM} = E - m\,r_0^4\;.
\eeq
Hence, combining (\ref{relad}) and (\ref{admtran}) we finally get 
\beq\label{enla}
\omega= E +  \half m \left(r_0^4 - 3 r_0^2 + {3\over 4}\right)\;.
\eeq
As a result, we may rewrite the semiclassical Hawking rate of D3-brane emission in terms of the canonical energy $E$ as
\beq\label{hawcan}
\Gamma_H (E) \propto e^{-\beta(E-E_0)}\;,
\eeq
where the threshold energy $E_0$ is given by
\beq\label{ecero}
E_0 = \mu_N - \half m \left(r_0^4 - 3 r_0^2 + {3\over 4}\right)= -m(r_0^4 - r_0^2)
\;,
\eeq
where we have used the formula $\mu_N = -\shalf m \left(r_0^2 - \shalf\right) \left(r_0^2 + \sthreeovertwo\right)$.
The critical energy $E_0$  is large and negative for large temperature, and vanishes at $r_0 =1$, corresponding to the special black hole isometric to pure AdS, with temperature $T= (2\pi)^{-1}$. On the other hand, $E_0$ becomes positive for
lower temperatures, approaching $m/4$ for the extremal black hole. The semiclassical form (\ref{hawcan}) is therefore valid for
$E >E_0$ and, as indicated previously, we assume that $\Gamma_H (E)$  loses the exponential suppression form  for
the  window $0<E< E_0$, which only opens up at low temperatures $T< (2\pi)^{-1}$.

\section{Decay rates}

\noindent

In this section we estimate the leading tunneling effects across the barriers that D-branes may find outside the horizon.  Since we have a very explicit description of the rigid motion in terms of the action (\ref{probecoor}), we use         
 the  WKB approximation  at fixed canonical energy $E$.   The  wave function is given, in the leading approximation, by  the {\it ansatz} 
\beq\label{wkba}
\Psi_E \propto \exp\left(-iEt + i\int^r p_{r'}\, dr' \right)
\;
\eeq
 where the radial canonical momentum is defined as $p_r = \pt L / \pt {\dot r}$. 
The probability of barrier penetration is given, with  exponential accuracy,  by 
\beq\label{probp}
\Gamma_E \sim \exp\left(-2\,{\rm Im}\, W(E)\right)\;,
\eeq
where 
$
W= I + E t
$
is the so-called  truncated action. The imaginary part in the classically forbidden region can be captured by the analytic continuation to the Euclidean signature $t=-i t_E$ and we find 
 ${\rm Im}\, W(E) \equiv W_E = I_E - E t_E$,  with   $I_E$  the Euclidean action
\beq\label{euclida}
I_E = m\int dt_E \left[r^3 \sqrt{ f(r)+ {1\over f(r)} \left( {d r \over dt_E}\right)^2} -\varepsilon \,(r^4 -r_0^4) \right]\;.
\eeq
Finally, Euclidean trajectories correspond  to motion in the effective problem
\beq\label{effeu}
\left({d r \over dt_E}\right)^2 = 2V_{\rm eff} (r)\;,
\eeq 
which is related to (\ref{effpot}) by a formal sign flip of the effective potential. 
Using this equation in the formula for the Euclidean action, we find the convenient expression
\beq\label{tune}
W_E = m\int_{r_-}^{r_+} {dr \over f(r)} \sqrt{r^6 \,f(r) - \left(r^4 -r_0^4  + {\varepsilon \,E\over m}\right)^2}\;.
\eeq
The integration domain is defined by the positivity of the square root argument, i.e. $W_E $  vanishes  when the turning points of the motion (\ref{effeu}) degenerate at a single radius $r_+ = r_- =r_s$, defined by  the largest solution $r_s >r_0$  of $V_{\rm eff} (r_s) = V'_{\rm eff} (r_s) =0$ (the horizon itself is always a solution of these equations). This happens at  the  sphaleron energy, $E_s = V_s (r_s)$,  which equals the value of the static potential evaluated at the static trajectory $r(t_E) = r_s$.

For  $\varepsilon =1$ and energies in the range $0 \leq E\leq E_s$ the tunneling exponent  $W_E $ governs the rate of barrier penetration for D-branes of energy $E$. For $E > E_s$ the barrier
disappears and $W_E (E > E_s ) =0$. On the other hand, for energies below the critical value $E < 0$ the barrier features a pole at $r_p = (r_0^4 -E /m)^{1/4}$,  in the physical region between the two turning points $r_0 < r_- < r_p < r_+$. Notice however that the expression (\ref{tune}) has no singular behavior  across the pole and has, in fact, quite a smooth dependence on the energy $E$.

\subsection{Pair creation}

\noindent 

For $E<0$, the Euclidean trajectories solving (\ref{effeu}) correspond to D-branes of energy $E$ propagating in the interval $r_p < r<r_+$ and to antiD-branes of energy $-E$ propagating in the interval $r_- < r<r_p$. Hence, we may interpret the solutions as the Euclidean description of the brane-antibrane
nucleation process, similar to Schwinger's description of an electric field decay by  $e^+ e^-$ emission \cite{schwinger}. The D-brane member of the pair emerges at $r_+$ and falls to infinity, whereas the antiD-brane emerges at $r_-$ and falls towards the black hole.  

The analogy with the Schwinger process can be made quite literal. The electron Lagrangian in the presence of a constant electric field in one dimension reads 
\beq\label{electron}
L_e = -m_e\sqrt{1-{\dot x}^2} + e\,{\cal E}\,x\;.
\eeq
Performing the same manipulations for this system we find an effective potential description
$
{\dot x}^2 + V_{\rm e} (x)=0\;,
$
with
\beq\label{elecpot}
V_{\rm e} (x) = {m_e^2 \over (E + e\,{\cal E}\,x)^2} - 1\;,
\eeq
which, with the exception of the warping effects, has essentially the same   structure as our brane potentials, including the occurrence of poles and the  rule that electrons propagate in the region $x>-E/ e{\cal E}$ and positrons do so in the region $x< -E /e{\cal E}$. The two turning points at $x_\pm = -E/e{\cal E} \pm m_e/e{\cal E}$ are interpreted semiclassically as the nucleation positions of  $e^+ e^-$ pairs in a  Schwinger decay process of the electric field. 
For the $e^+ e^-$ potential (\ref{elecpot}), equation (\ref{tune}) gives the expected exponent $2W_E =
2\int_{x_-}^{x_+} dx \sqrt{m_e^2 - (e{\cal E} x)^2} = \pi m_e^2/ e {\cal E}$ controlling the Schwinger effect amplitude.

We can estimate the WKB factor (\ref{tune}) for very large temperatures, $r_0 \gg 1$ by noticing that the turning points, defined by the vanishing of the square root in the integrand (\ref{tune}) are given by
\beq\label{tup}
r_- \approx {r_p^2 \over (2r_p^4 - r_0^4)^{1/4}}\;, \qquad r_+ \approx (2r_p^4 - r_0^4)^{1/2}\;,
\eeq
in this limit. In these expressions, we denote $r_p = (r_0^4 - E/m)^{1/4}$ the pole position, and we recall that
$E<0$ in our conventions. At high temperatures, $r_0 \gg 1$, we have $r_+ \gg r_-$, and the integral in (\ref{tune}) is dominated by the high endpoint, so that we have
\beq\label{weap}
W_E \approx m (2r_p^4 -r_0^4) \int_0^1 dx \,\sqrt{1-x^2} \approx {\pi \over 4} m \,(r_0^4 - 2E/m)\;.
\eeq
Using the asymptotic relation $r_0 \sim \pi T \gg1$, we find an amplitude $\exp(-2W_E)$ with
\beq\label{sex}
2W_E \approx {\pi^3 \over 4} N\,V\,T^4 + \pi |E|\;,
\eeq
where the second term is only to be kept when it dominates over the first, i.e. we have an amplitude
of order $\exp(-\pi |E|)$ for energies very large and negative, $-E \gg 1$, so that $r_p \gg r_0$. The complete Schwinger amplitude must be integrated over all negative energies or, equivalently over all pole locations $r_p \geq r_0$. In view of (\ref{sex}), this integral is dominated by the endpoint at zero energy, $E=0$, which yields a total Schwinger amplitude at high temperatures
\beq\label{endps}
\Gamma_{\rm S} \sim \exp\left(-{\pi^3 \over 4} N\,V\, T^4 \right)\;,
\eeq
as always, in units $\ell=1$.

\subsection{Thermal emission of D-branes}

\noindent

For $E>0$ the Schwinger process does not take place, as the pole migrates behind the horizon. Instead, we have
a more standard process of tunneling across the potential barrier, with the initial state being fed by the Hawking radiation law. The barrier is only effective for energies in the window $0<E<E_s$. Our analysis of the previous section showed that the correct rate of D3-branes impinging on the barrier is given by 
\beq\label{hawrotr}
\Gamma_H (E) \sim e^{-\beta(E-E_0)} \;,
\eeq
provided $E>E_0 = - m (r_0^4 - r_0^2)$. In the window $0<E<E_0$, which only opens up for $T< (2\pi)^{-1}$,
we assume that the decay rate is controlled by the temperature $\Gamma_H (E) \sim T$ and it is not expected to depend exponentially on the energy of the brane or the charge $N$ of the black hole. 

We see that for all temperatures for which $E_s > E_0$ the metastability of the black hole is protected either by
the exponential suppression of the Hawking rate or by the presence of the external barrier. This is the case at large temperatures. On the other hand, at low temperatures $E_s$ vanishes, whereas $E_0$ approaches the positive value $m/4$, resulting in the opening of a window, $E_s < E < E_0$, for which the D-branes escape without exponential suppression from neither the Hawking rate nor the grey body factor. The critical temperature at which metastability is lost is of order $T_c \approx 0.12$ in units $\ell=1$, and can be found by solving numerically the equation
$E_0 = E_s$. 

The breakdown of metastability for $T<T_c$ can be suggestively interpreted as the strong-coupling counterpart of the low-$T$ thermal mass being dominated by the tachyonic mass at weak coupling.

At $T>T_c$ the black hole is indeed metastable, and the total emission rate can be approximated by the convolution 
\beq\label{barr}
\Gamma_{\rm thermal}  \sim \int_{0}^{\infty} dE \,\Gamma_H (E) \;e^{ - 2W_E} \;.
\eeq
Using the exponential form and performing a standard saddle-point evaluation (see for example \cite{affleck}) we find 
$$
 e^{\beta E_0} \int dE\,e^{-\beta E -2W_E } \sim e^{\beta E_0 } e^{- I_E (\beta)}\;,
$$
where $I_E (\beta)$ is the value of the Euclidean action at a periodic trajectory solving 
(\ref{effeu})
with period $\beta$. The period of such thermal bounces  is larger the larger is the barrier. The minimum period is attained when the barrier disappears, and is given by $2\pi / \Omega_s$, where $\Omega_s^2 = V_{\rm eff}'' (r_s)$
is the second derivative of the effective potential at the sphaleron value of the energy $E_s$. Hence, for
$\beta < 2\pi / \Omega_s$ there are no possible bounces with that period and the rate is dominated by the {\it static} solution, corresponding to the constant  trajectory $r(t_E)= r_s$
for the potential at energy $E_s$. Computing the action for this solution one finds
\beq\label{spaac}
e^{-I_E ({\rm sphaleron})} = e^{-\beta V_s (r_s)}  = e^{-\beta \,E_s}
\;.
\eeq
 At large temperatures the sphaleron frequency $\Omega_s = {\cal O}(1)$, so that for any large temperature 
 $T\gg 1$ in units of the AdS radius, the rate will be given
by the sphaleron approximation 
\beq\label{sphasl}
\Gamma_{\rm thermal} \big |_{ T\gg 1} \sim e^{-\beta(E_s - E_0)} \sim \exp\left(-{\pi^2 \over 4} NVT^3\right)\;.
\eeq
Incidentally, this has the same order of magnitude, within exponential accuracy, as the high-energy endpoint contribution, corresponding to the high-energy Hawking tail without grey-body suppression, $\int_{E_s}^\infty dE\, e^{-\beta (E-E_0)} \sim e^{-\beta(E_s -E_0)}$. 

   \begin{figure}
  \label{nump}
  \begin{center}
    \epsfig{file=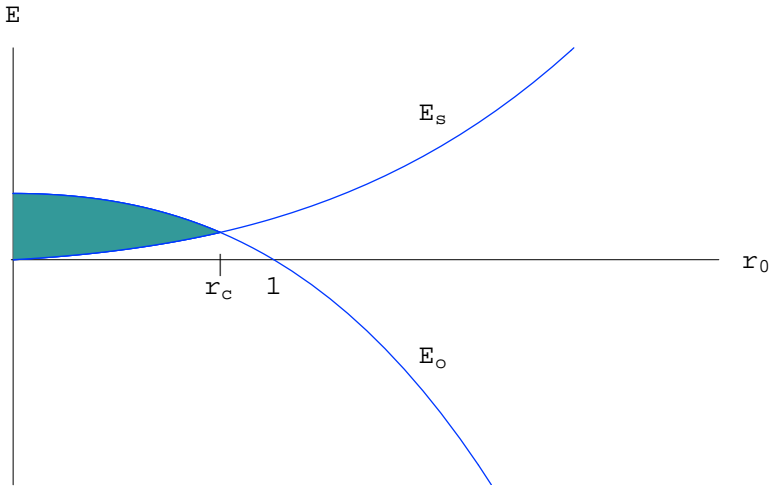, width= 10.5cm}
   \caption{ \small  Comparison of the threshold energy $E_0$ and the sphaleron energy $E_s$ as a function of temperature, or rather horizon radius in this picture. The shaded region, at horizon radii $r\leq r_c \approx 0.93 $, represents the window of D-brane energies $E_s < E< E_0$  for which metastability is lost.}
  \end{center}
\end{figure}

\section{Conclusions}

\noindent

We have analyzed the metastability of the $SU(N)$ symmetric state of  ${\cal N} =4$ SYM theory
on a compact  three-dimensional hyperboloid, using the AdS bulk description of the system at large $N$ and large  values of the 't Hooft coupling, $\lambda = g_{\rm YM}^2 N \gg 1$. The zero-mode energy of the Higgs fields is  unbounded below in this theory, whereas one expects the $SU(N)$-symmetric state at vanishing values of the Higgs fields to be locally stable at finite temperature. We test this expectation in the supergravity approximation and estimate the corresponding rates for the decay of the $SU(N)$-symmetric metastable state into $SU(N-1)$, times a runaway $U(1)$ factor.

Our main results are as follows. At large temperatures $T\gg (2\pi\ell)^{-1}$ the thermal states are metastable, with an exponentially suppressed decay rate of order
\beq\label{finlarge}
\Gamma_{\rm thermal} \big |_{2\pi\ell\, T \gg 1} \sim \exp \left(-{\pi^2 \over 4} NVT^3\right)\;,
\eeq
and dominated by thermal excitation over the `grey-body' barrier, although a competing effect with a smaller coefficient also manifests itself by quantum nucleation of brane-antibrane pairs, in an analog of Schwinger's process, 
\beq\label{finsw}
\Gamma_{\rm S} \big |_{2\pi \ell\, T \gg 1} \sim \exp\left(-{\pi^2 \over 4} NVT^3 (\pi \ell T)\right)\;.
\eeq
We see that the pair production effect is exponentially suppressed with respect to the thermal excitation over the barrier (at large temperatures), by an extra relative factor of $\pi T \ell \gg 1$ in the exponent. 

A critical temperature exists,  $T_c \approx (8.3\, \ell)^{-1} < (2\pi\ell)^{-1}$,  below which the thermal emission process loses its exponential suppression with the RR charge $N$, becoming a much faster powerlike
rate which renders the low-temperature black holes much less stable than their large-temperature counterparts.
It would be interesting to improve the analysis of this low-temperature `superradiant' regime to obtain an estimate of the powerlike behavior. The main difficulty in this task lies in the corrections to the Hawking rate of D-branes, rather than the
next-to-leading corrections in the calculation of barrier penetration. It is interesting that the special topological black hole at $r_0 =1$, locally isometric to pure AdS, falls into the metastable domain according to our results.

We conclude that a hot black hole decays extremely slowly until the temperature drops below the critical value, set by the AdS curvature, when the process accelerates to a rate of order $T$. Eventually the value of $N$ and $\lambda = g_{\rm YM}^2 N$ gradually decrease in the interior geometry that is left over. When the `interior' value of the 't Hooft coupling reaches ${\cal O}(1)$, a matching to a weakly-coupled description is required, since the strong curvature prevents the use of geometrical methods (see \cite{lowe} for preliminary results in this direction). 
It would be interesting to study whether the lack of metastability at low temperatures that we found here extends to the perturbative Yang--Mills domain, by an explicit computation of the thermal effective potential on the compact hyperboloid. 
 Such an analysis would surely shed light on the use of these  backgrounds in the program spelled out \cite{horo}.

\subsubsection*{Acknowledgements}

\noindent

We are indebted to R. Emparan and especially E. Rabinovici for discussions. 
 This work was partially supported by MEC and FEDER under grants FPA2006-05485 and FPA2009-07908,  the Spanish
Consolider-Ingenio 2010 Programme CPAN (CSD2007-00042) and  Comunidad Aut\'onoma de Madrid under grants HEPHACOS P-ESP-00346 and HEPHACOS S2009/ESP-1473. J.M.M. is supported by a FPU fellowship  AP2008-00149
from MICINN.

%%%%%%%%%%%%%%%%%%%%%%


\begin{thebibliography}{77}


 \bibitem %[toporefs]
{toporefs}
J.P.~Lemos,
``Cylindrical black hole in general relativity,"
Phys. Lett. {\bf B353}, 46 (1994)
gr-qc/9404041;

J.P.~Lemos,
``Two-dimensional black holes and planar general relativity,"
Class. Quant. Grav. {\bf 12}, 1081 (1995)
gr-qc/9407024;

S.~Aminneborg, I.~Bengtsson, S.~Holst and P.~Peld{\'a}n,
``Making anti-de Sitter black holes,"
Class. Quant. Grav. {\bf 13}, 2707 (1996)
gr-qc/9604005;

R.B.~Mann,
``Pair production of topological anti-de Sitter black holes,"
Class. Quant. Grav. {\bf 14}, L109 (1997)
gr-qc/9607071;

D.~Brill, J.~Louko and P.~Peld{\'a}n, 
Phys. Rev. {\bf D56}, 3600 (1997)
gr-qc/9705012;

L.~Vanzo,
``Black holes with unusual topology,"
Phys. Rev. {\bf D56}, 6475 (1997)
gr-qc/9705004;


R.B.~Mann, 
``Topological Black Holes: Outside Looking In,"
gr-qc/9709039.


\bibitem{roberto}

R.~Emparan,
  ``AdS membranes wrapped on surfaces of arbitrary genus,''
  Phys.\ Lett.\  B {\bf 432}, 74 (1998)
  [arXiv:hep-th/9804031].
  
  R.~Emparan,
  ``AdS/CFT duals of topological black holes and the entropy of  zero-energy
  states,''
  JHEP {\bf 9906}, 036 (1999)
  [arXiv:hep-th/9906040].

  

 



\bibitem{topoantiguo}

D.~Birmingham,
  ``Topological black holes in anti-de Sitter space,''
  Class.\ Quant.\ Grav.\  {\bf 16}, 1197 (1999)
  [arXiv:hep-th/9808032].

D.~Birmingham and M.~Rinaldi,
  ``Brane world in a topological black hole bulk,''
  Mod.\ Phys.\ Lett.\  A {\bf 16}, 1887 (2001)
  [arXiv:hep-th/0106237].
  
    
  D.~Birmingham and S.~Mokhtari,
  ``Stability of Topological Black Holes,''
  Phys.\ Rev.\  D {\bf 76}, 124039 (2007)
  [arXiv:0709.2388 [hep-th]].
  
  
   
  \bibitem{trodden}


N.~Kaloper, J.~March-Russell, G.~D.~Starkman and M.~Trodden,
  ``Compact hyperbolic extra dimensions: Branes, Kaluza-Klein modes and
  cosmology,''
  Phys.\ Rev.\ Lett.\  {\bf 85}, 928 (2000)
  [arXiv:hep-ph/0002001]. 
  
  Mark Trodden,''Diluting Gravity with Compact Hyperboloids``, arXiv:hep-th/0010032

 \bibitem{evauno}
  J.~McGreevy, E.~Silverstein and D.~Starr,
  ``New dimensions for wound strings: The modular transformation of geometry to
  topology,''
  Phys.\ Rev.\  D {\bf 75} (2007) 044025
  [arXiv:hep-th/0612121].
  
  D.~R.~Green, A.~Lawrence, J.~McGreevy, D.~R.~Morrison and E.~Silverstein,
  ``Dimensional Duality,''
  Phys.\ Rev.\  D {\bf 76} (2007) 066004
  [arXiv:0705.0550 [hep-th]].

\bibitem{frag}
J.~M.~Maldacena, J.~Michelson and A.~Strominger,
  ``Anti-de Sitter fragmentation,''
  JHEP {\bf 9902} (1999) 011
  [arXiv:hep-th/9812073].
  
  
\bibitem{seiwit}
N.~Seiberg and E.~Witten,
  ``The D1/D5 system and singular CFT,''
  JHEP {\bf 9904}, 017 (1999)
  [arXiv:hep-th/9903224].

\bibitem{buchel}
A.~Buchel,
  ``Gauge theories on hyperbolic spaces and dual wormhole instabilities,''
  Phys.\ Rev.\  D {\bf 70}, 066004 (2004)
  [arXiv:hep-th/0402174].

\bibitem{rabadan}
M.~Kleban, M.~Porrati and R.~Rabadan,
  ``Stability in asymptotically AdS spaces,''
  JHEP {\bf 0508}, 016 (2005)
  [arXiv:hep-th/0409242].
  
\bibitem{horo}
  G.~Horowitz, A.~Lawrence and E.~Silverstein,
  ``Insightful D-branes,''
  JHEP {\bf 0907}, 057 (2009)
  [arXiv:0904.3922 [hep-th]].
  %%CITATION = JHEPA,0907,057;%%


\bibitem{eva}

L.~Kofman, A.~D.~Linde, X.~Liu, A.~Maloney, L.~McAllister and E.~Silverstein,
  ``Beauty is attractive: Moduli trapping at enhanced symmetry points,''
  JHEP {\bf 0405}, 030 (2004)
  [arXiv:hep-th/0403001].

  
\bibitem{kiritsis}
  E.~Kiritsis,
  ``Supergravity, D-brane probes and thermal super Yang-Mills:  A comparison,''
  JHEP {\bf 9910}, 010 (1999)
  [arXiv:hep-th/9906206].
 
  E.~Kiritsis and T.~R.~Taylor,
  ``Thermodynamics of D-brane probes,''
  arXiv:hep-th/9906048.




  
   
  


  

  

\bibitem{wil}

M.~K.~Parikh and F.~Wilczek,
  ``Hawking radiation as tunneling,''
  Phys.\ Rev.\ Lett.\  {\bf 85}, 5042 (2000)
  [arXiv:hep-th/9907001]. 
  
  P.~Kraus and F.~Wilczek,
  ``A Simple stationary line element for the Schwarzschild Geometry, and some
  applications,''
  arXiv:gr-qc/9406042. 
  
  P.~Kraus and F.~Wilczek,
  ``Self-Interaction Correction to Black Hole Radiance,''
  Nucl.\ Phys.\  B {\bf 433}, 403 (1995)
  [arXiv:gr-qc/9408003].
  
  P.~Kraus and F.~Wilczek,
  ``Effect Of Selfinteraction On Charged Black Hole Radiance,''
  Nucl.\ Phys.\  B {\bf 437}, 231 (1995)
  [arXiv:hep-th/9411219].
  
  \bibitem{israel}
  W.~Israel,
  ``Singular hypersurfaces and thin shells in general relativity,''
  Nuovo Cim.\  B {\bf 44S10}, 1 (1966)
  [Erratum-ibid.\  B {\bf 48}, 463 (1967\ NUCIA,B44,1.1966)].
  
  \bibitem{junct}
  S.~K.~Blau, E.~I.~Guendelman and A.~H.~Guth,
  ``The Dynamics of False Vacuum Bubbles,''
  Phys.\ Rev.\  D {\bf 35}, 1747 (1987).
  %%CITATION = PHRVA,D35,1747;%%

 V.~A.~Berezin, V.~A.~Kuzmin and I.~I.~Tkachev,
  ``Thin Wall Vacuum Domains Evolution,''
  Phys.\ Lett.\  B {\bf 120}, 91 (1983).
  
  V.~A.~Berezin, V.~A.~Kuzmin and I.~I.~Tkachev,
  ``Dynamics of Bubbles in General Relativity,''
  Phys.\ Rev.\  D {\bf 36}, 2919 (1987).
  
 
  A.~Aurilia, G.~Denardo, F.~Legovini and E.~Spallucci,
  ``Vacuum Tension Effects On The Evolution Of Domain Walls In The Early
  Universe,''
  Nucl.\ Phys.\  B {\bf 252}, 523 (1985).
  
   G.~L.~Alberghi, D.~A.~Lowe and M.~Trodden,
  ``Charged false vacuum bubbles and the AdS/CFT correspondence,''
  JHEP {\bf 9907}, 020 (1999)
  [arXiv:hep-th/9906047].
   
B.~Freivogel, V.~E.~Hubeny, A.~Maloney, R.~C.~Myers, M.~Rangamani and S.~Shenker,
  ``Inflation in AdS/CFT,''
  JHEP {\bf 0603}, 007 (2006)
  [arXiv:hep-th/0510046].

\bibitem{us}
J.~L.~F.~Barbon and E.~Rabinovici,
  ``Holography of AdS vacuum bubbles,''
  arXiv:1003.4966 [Unknown].
  



\bibitem{affleck}
 I.~Affleck,
  ``Quantum Statistical Metastability,''
  Phys.\ Rev.\ Lett.\  {\bf 46} (1981) 388.
  




\bibitem{schwinger}
J.~S.~Schwinger,
  ``On gauge invariance and vacuum polarization,''
  Phys.\ Rev.\  {\bf 82} (1951) 664.

\bibitem{lowe}
N.~Iizuka, D.~N.~Kabat, G.~Lifschytz and D.~A.~Lowe,
  ``Probing black holes in non-perturbative gauge theory,''
  Phys.\ Rev.\  D {\bf 65} (2002) 024012
  [arXiv:hep-th/0108006].

  \end{thebibliography}
\end{document}